\documentclass[12pt]{article}
\usepackage{amssymb}
\usepackage{amsfonts,amssymb,amsmath,amsthm}
\usepackage{epsfig}
\usepackage{slashed}
\usepackage{graphicx}
\usepackage{mathrsfs}
\usepackage{bbm}
\def\and{{\rm and}}

\def\d{\delta}

\def\om{\omega}
\def\p{\partial}

\def\O{\Omega}

\begin{document}
\renewcommand{\thefootnote}{\fnsymbol{footnote}}
\begin{titlepage}
\begin{flushright}
USTC-ICTS-10-21
\end{flushright}

\vspace{10mm}
\begin{center}
{\Large\bf Massive charged particle's tunneling from spherical charged black hole}
\vspace{16mm}

{{\large Yan-Gang Miao${}^{1,2,}$\footnote{\em E-mail: miaoyg@nankai.edu.cn}, Zhao Xue${}^{1,}$\footnote{\em E-mail: illidanpimbury@mail.nankai.edu.cn}
and Shao-Jun Zhang${}^{1,}$}\footnote{\em E-mail: sjzhang@mail.nankai.edu.cn}\\

\vspace{6mm}
${}^{1}${\em School of Physics, Nankai University, Tianjin 300071, \\
People's Republic of China}

\vspace{3mm}
${}^{2}${\em Interdisciplinary Center for Theoretical Study,\\
University of Science and Technology of China, Hefei, Anhui 230026,\\
People's Republic of China}}

\end{center}

\vspace{10mm}
\centerline{{\bf{Abstract}}}
\vspace{6mm}
%The Hawking radiation can be viewed as a tunneling process and computed by the WKB method semiclassically. In this note,
We generalize the Parikh-Wilczek scheme to the tunneling of a massive charged particle from a general spherical charged black hole.
We obtain that the tunneling probability depends on the energy, the mass and the charge of the particle.
%which indicates that the exact emission spectrum is not precisely thermal. Through equating the tunneling probability to the Boltzmann factor,
In particular, the modified Hawking temperature is related to the charge.
Only at the leading order approximation can the standard Hawking temperature be reproduced.
We take the Reissner-Nordstr\"{o}m black hole as an example to clarify our points of view,
and find that the accumulation of Hawking radiation makes it approach an
extreme black hole.
%if the tunneling particle has a mass, the limits of integration should be treated carefully. This leads to a correction to the emission spectrum.
%And the charge of the particle also participates in the tunneling rate. We can see that the spectrum is no longer thermal precisely.
%To the first order approximation, we show that in the special case--the Reissner-Nordstr\"{o}m black hole,
%the Hawking temperature seems effective according to the charge of the tunneling particle.
\vskip 20pt
PACS Number(s): 04.70.Dy, 03.65.Sq, 04.62.+v
\vskip 10pt
Keywords:
Black hole, quantum tunneling, Hawking temperature
\end{titlepage}

\newpage
\renewcommand{\thefootnote}{\arabic{footnote}}
\setcounter{footnote}{0}
\setcounter{page}{2}
\pagenumbering{arabic}

\section{Introduction}
Whether a black hole can radiate puzzled physicists for a long time after the concept of black holes was put forward.
%It relates to the entropy difficult.
In 1974, Hawking proved~\cite{Hawking} that black holes are not really black, and that
they radiate energy continuously and have an emission spectrum of a black body. Nowadays, the radiation of black holes is called ``Hawking radiation".
%And quantum tunneling was viewed as the heuristic picture of the radiation naturally.
Although Hawking proposed that the radiation of black holes can be viewed as tunneling, he
calculated the emission spectrum in light of quantum field theory in curved spacetime without following the tunneling picture.
%Later on Hawking developed methods to compute the radiation based on quantum field theory~\cite{Hawking75}.
%The result presented that black hole has an emission spectrum of black body.
%But the derivation corresponding with the heuristic physical picture wasn't proposed for a long time.

In 1999, Parikh and Wilczek presented~\cite{Wilczek99} a concise and direct derivation of the Hawking radiation as a tunneling process.
They suggested that the barrier is created by the tunneling particle itself, and
utilized the Wentzel-Kramers-Brillouin (WKB) method to compute the tunneling rate semiclassically. That is, the WKB expansion was made to the first order,
and the Hawking temperature was then obtained in terms of equating the tunneling rate to the Boltzmann factor.
The key point of their scheme is to find out a coordinate system in which the metric is well-behaved for calculations at
the event horizon.
%The imaginary part of the action of the tunneling particle is needed to calculate the tunneling
%rate following the standard arguments.
%The integrand involves the null geodesic, which is the trajectory of the massless tunneling particles.
%And pole of the equation of the null geodesic can afford the imaginary part of the action.
Following the scheme, they derived the emission spectrum of massless and uncharged particles from spherically symmetric black holes,
such as the Schwarzschild black hole and Reissner-Nordstr\"{o}m black hole. Their results show that although the radiation is not precisely thermal,
the leading order is consistent with the standard Hawking radiation, i.e. the black body spectrum.
Therefore, Parikh and Wilczek provided a reasonable scheme for the study of the Hawking radiation as a tunneling process.
%they're thought to be on the right way.
However, they just considered the massless and uncharged particle's tunneling. Later, an attempt was made~\cite{Zheng Zhao JHEP}
for the tunneling of a massive and charged particle, in which the
main idea is to introduce the concepts of phase velocity and group velocity
and to equate the radial velocity of the tunneling particle to the phase velocity.

%After that, plenty of papers presented to discuss different kind of black holes and different tunneling processes,
%here we list some~\cite{Zheng Zhao NPB,Zheng Zhao JHEP,Bin Chen,Mehdipour}.
%However when some of them ~\cite{Zheng Zhao NPB,Mehdipour}discuss tunneling particles with mass,
%Furthermore, the bound of integral was not treated seriously,
%which has an important influence of the spectrum as we will see in this paper.

In addition, based on the Hamilton-Jacobi equation Srinivasan
et al. suggested~\cite{Srinivasan} an alternative approach to study the Hawking radiation as a tunneling process.
As this approach is not involved in the trajectory problem,
it can be applied to the tunneling particles with mass or charge in a natural way.
The approach was later developed to extremal and rotating black holes~\cite{Angheben:2005}, to Taub-NUT black holes~\cite{Kerner} and
to quantum tunneling beyond semiclassical approximation~\cite{Banerjee}.

Our motivation is straightforward in this note, i.e. to generalize the Parikh-Wilczek scheme~\cite{Wilczek99}
to the tunneling process of massive and charged particles from general spherical charged black holes.
Different from the ways given by ref.~\cite{Srinivasan} and ref.~\cite{Zheng Zhao JHEP},
we neither circumvent the trajectory problem nor adopt the concepts of phase velocity and group velocity.
What we want to do is to follow exactly the idea of Parikh and Wilczek and
to derive strictly the equation of radial motion for the massive charged tunneling particle.
We note that in this situation the trajectory of the particle is time-like due to its mass,
and is not geodesic anymore due to the electromagnetic interaction between the particle and black hole.
In particular, our result, different from that of
ref.~\cite{Zheng Zhao JHEP}, can on the one hand recover that of Parikh and Wilczek in the limit of zero charge, and on the other hand
ensure the dispersion relation of the tunneling particle, which coincides with the basic requirement that
the observer at infinity can detect the particle only on shell.

This note is organized as follows. In section 2, we derive the equation of radial motion for the tunneling particle with mass and charge.
%But it still can be computed.
In section 3, we calculate the imaginary part of the action of the tunneling particle
and then give a general expression of the modified Hawking temperature.
In section 4 we take the Reissner-Nordstr\"{o}m black hole as an example for the application of our generalization of the Parikh-Wilczek scheme.
We obtain the corresponding tunneling
rate and the Hawking temperature, which can clarify explicitly our points of view. Finally, we make a conclusion in section 5.
%to get an explicit form of the emission spectrum with our method. It is modified and not precisely
%thermal. We can also treat the leading order as thermal,
%we will see the influence of mass and charge of particles. In some limit, we can still regain Hawking's result.

\section{Radial motion of massive charged particle}
A general static spherical charged black hole with mass $M$ and charge $Q$ is depicted by the following metric and electromagnetic potential,
\begin{eqnarray}
d s^2 &=& - f(r) d t^2+ g^{-1}(r) d r^2 + r^2 d \O^2,\label{metric}\\
A_\mu &=& h(r) \d_\mu^0,\label{potential}
\end{eqnarray}
where $f(r)$, $g(r)$ and $h(r)$ are functions of the parameters: $M$, $Q$ and $r$.
As was done in ref.~\cite{Wilczek99}, it is necessary to work in the Painlev\'{e} coordinate in order to calculate the Hawking temperature
via particle's tunneling.
The Painlev\'{e} coordinate can be acquired by the transformation~\cite{Painleve:1921},
\begin{equation}
d t_p = d t + \sqrt{\frac{1 - g(r)}{f(r) g(r)}} d r,
\end{equation}
where $t_p$ denotes the Painlev\'{e} time. The resulting Painlev\'{e} line element and electromagnetic potential take the forms,
\begin{eqnarray}
d s^2 &=& - f(r) d t_p^2 +  d r^2 + 2 f(r) \sqrt{\frac{1 - g(r)}{f(r) g(r)}} d t_p d r + r^2 d \O^2,\label{metric1}\\
A_\mu &=& h(r) \d_\mu^0 - h(r) \sqrt{\frac{1 - g(r)}{f(r) g(r)}} \d_\mu^1.\label{potential1}
\end{eqnarray}

The motion of a particle with mass $m$ and charge $q$ in such a spacetime can be described by the following
Lagrangian\footnote{The Lagrangian can be obtained from the covariant equation of motion,
\begin{equation*}
\frac{d^2 x^\mu}{d \tau^2} + \Gamma^\mu_{\nu\rho} \frac{d x^\nu}{d \tau} \frac{d x^\rho}{d \tau} = \frac{q}{m} F^\mu_{~\nu} \frac{d x^\nu}{d \tau}.
\end{equation*}},
\begin{equation}
\mathcal{L} = \frac{m}{2} g_{\mu\nu} \frac{d x^\mu}{d \tau} \frac{d x^\nu}{d \tau} + q A_\mu \frac{d x^\mu}{d \tau},
\end{equation}
where $\tau$ is chosen to be the proper time. Substituting the solution eqs.~(\ref{metric1}) and (\ref{potential1}) into the above formula,
the Lagrangian can be rewritten as
\begin{equation}
\mathcal{L} = \frac{m}{2} \left[ - f(r) \dot{t_p}^2 + \dot{r}^2 + 2 f(r) \sqrt{\frac{1 - g(r)}{f(r) g(r)}} \dot{t_p} \dot{r}\right] + q h(r) \dot{t_p}
- q h(r) \sqrt{\frac{1 - g(r)}{f(r) g(r)}} \dot{r},\label{Lagrangian}
\end{equation}
where the dot stands for the derivative with respect to $\tau$, i.e. $\dot{x}^\mu \equiv \frac{d x^\mu}{d \tau}$.
By using the Lagrangian (eq.~(\ref{Lagrangian}))
and its corresponding Euler-Lagrange equation, we can derive the following equation of motion with respect to coordinate $t_p$,
\begin{equation}
- p_{t_p} \equiv - \frac{\p \mathcal{L}}{\p \dot{t_p}} = m f(r) \dot{t_p} - m f(r) \sqrt{\frac{1 - g(r)}{f(r) g(r)}} \dot{r} - q h(r)
\equiv \omega = {\rm const}.,\label{motioneqn1}
\end{equation}
where the minus sign before $p_{t_p}$ is present due to the positivity of the energy $\omega$ of the tunneling particle.
For a time-like trajectory, we also have
\begin{equation}
g_{\mu\nu} \dot{x}^\mu \dot{x}^\nu = -1,
\end{equation}
which can be explicitly written as
\begin{equation}
- f(r) \dot{t_p}^2 + \dot{r}^2 + 2 f(r) \sqrt{\frac{1 - g(r)}{f(r) g(r)}} \dot{t_p} \dot{r} = -1.\label{motioneqn2}
\end{equation}
With eqs.~(\ref{motioneqn1}) and (\ref{motioneqn2}), we can obtain
\begin{eqnarray}
\dot{r} &=& \pm \frac{1}{m} \sqrt{\frac{g(r)}{f(r)} \left(\omega + q h(r)\right)^2 - m^2 g(r)},\label{motioneqn3}\\
\dot{t_p} &=& \frac{1}{m f(r)} \left[\left(\omega + q h(r)\right) \pm \sqrt{1 - g(r)} \sqrt{\left(\omega + q h(r)\right)^2
- m^2 f(r)}\right],\label{motioneqn4}
\end{eqnarray}
where the upper (lower) sign in eqs.~(\ref{motioneqn3}) and (\ref{motioneqn4}) corresponds to outgoing (ingoing) trajectories.
As a consequence, we can determine the radial motion with respect to the Painlev\'{e} time $t_p$ by using eqs.~(\ref{motioneqn3}) and (\ref{motioneqn4}),
\begin{eqnarray}
\frac{d r}{d t_p} &=& \dot{r} \dot{t_p}^{-1}\nonumber\\
&=&\pm f(r) \frac{\sqrt{\frac{g(r)}{f(r)} \left(\omega + q h(r)\right)^2 - m^2 g(r)}}{(\omega + q h(r)) \pm \sqrt{1 - g(r)}
\sqrt{\left(\omega + q h(r)\right)^2 - m^2 f(r)}}.\label{radialmotion}
\end{eqnarray}

\section{Hawking radiation via particle's tunneling}
To derive the Hawking temperature in the tunneling picture, one has to calculate the imaginary part of the action of the tunneling particle.
In ref.~\cite{Wilczek99}, this imaginary part for the tunneling particle which crosses the horizon outwards
from $r_{in}$ to $r_{out}$ is defined as
\begin{eqnarray}
{\rm Im} S &\equiv& {\rm Im}\int^{r_{out}}_{r_{in}} p_r d r = {\rm Im} \int^{r_{out}}_{r_{in}} \int^{p_r}_0 d \tilde{p}_r d r
= {\rm Im} \int^{r_{out}}_{r_{in}} \int^{\omega}_{m} \frac{d H}{\frac{d r}{d t_p}} d r\nonumber\\
&=& - {\rm Im} \int^{r_{out}}_{r_{in}} \int^{\omega}_{m} \frac{d \tilde{\omega}}{\frac{d r}{d t_p}} d r,\label{imaction}
\end{eqnarray}
where the Hamilton equation $\frac{d r}{d t_p} = \frac{d H}{d \tilde{p}_r}$ has been utilized,
and the minus sign appears due to the relation~\cite{Wilczek99}:
$H=M-\tilde{\omega}$. Note that the tunneling particle is massive in our case,
thus the lower limit of energy integral is no longer zero but the particle's mass.
This consideration is
indispensable\footnote{If the tunneling particle arrives at infinity and is detected by an observer,
the trajectory eq.~(\ref{motioneqn3}) must be valid there.
For an approximately flat metric, we have $f(\infty)\sim g(\infty)\sim1$, and the electromagnetic potential
$A_\mu(\infty)\sim0$ which leads to $h(\infty)\sim0$ due to the static electromagnetic source, such as the Reissner-Nordstr\"{o}m black hole
we shall discuss in the next section. As a result, we deduce $\omega>m$ from eq.~(\ref{motioneqn3}), which coincides with the dispersion relation
of the tunneling particle at infinity.} for a massive tunneling particle
and we shall see that it will lead to a nontrivial effect,
i.e. a shift with respect to the energy on the emission spectrum (see the last paragraph in section 4 for the details).
Moreover, we emphasize that the insertion of the lower limit of mass on the integral in eq.~(\ref{imaction})
does not affect the contour integration\footnote{The authors would like to thank
the anonymous referee for suggesting a discussion on how the lower limit of mass affects the contour integration here,
and suggesting an expression for eq.~(\ref{RNTH})
in power of $q$ and an explanation about the absence of effects on the deformed Hawking temperature caused by the particle's mass in the next section.}.
The reason is obvious because the contour integration is only associated with the integration of $dr$ while the lower limit of mass appears
in the integration of $d\tilde\omega$. The concrete processing is as follows.
We use the Parikh-Wilczek's method~\cite{Wilczek99} which is different from the Hamilton-Jacobi method adopted in ref.~\cite{SKW},
and in light of the former
we exchange the order of the integration in eq.~(\ref{imaction}).
When we do the integration of $dr$, the lower energy integral limit existing in the integration of $d\tilde\omega$ certainly has no effects. Next,
when we do the integration of $d\tilde\omega$,
the lower energy limit then contributes a constant shift as appeared in eq.~(\ref{action}) up to the first order of $\tilde\omega$
(see eq.~(\ref{expansion})).
As a consequence, the integration in eq.~(\ref{imaction}) is not
sensitive to the limit $m \rightarrow 0$, i.e. one can take this limit before or after integration.

The horizon $r_h$ is given by $f(r_h) = g(r_h) = 0$ and is a function of $M$ and $Q$: $r_h = r_h (M, Q)$.
Near the horizon one can expand $f(r)$ and $g(r)$ with respect to the radial coordinate $r$ and to the horizon $r_h$ as the center,
\begin{eqnarray}
f(r) &=& f'(r_h) (r-r_h) + \mathcal{O}\left((r-r_h)\right),\nonumber\\
g(r) &=& g'(r_h) (r-r_h) + \mathcal{O}\left((r-r_h)\right).
\end{eqnarray}
Substituting the above equation into eq.~(\ref{radialmotion}), $\frac{d r}{d t_p}$ can be approximately expressed as
\begin{equation}
\frac{d r}{d t_p} \simeq \frac{\sqrt{f'(r_h) g'(r_h) \left(\omega + q h(r)\right)^2
- m^2 {f^{\prime}}^2(r_h) g'(r_h) (r-r_h)}}{\left(\omega + q h(r)\right) + \sqrt{1 - g'(r_h) (r-r_h)} \sqrt{\left(\omega
+ q h(r)\right)^2 - m^2 f'(r_h) (r-r_h)}} (r-r_h). \label{radialmotion1}
\end{equation}
Considering the self-gravitation of the tunneling particle, eq.~(\ref{radialmotion1}) should be modified by the replacement of
$M \rightarrow M-\omega$ and $Q \rightarrow Q-q$. Substituting eq.~(\ref{radialmotion1}) into eq.~(\ref{imaction})
and switching the order of integration
for doing the $r$ integral first, we see that there is a pole at $r=r_h$.
Completing the $r$ integral by deforming the contour around the pole and using the residue method, we have
\begin{equation}
{\rm Im} S = \pi \int^{\omega}_{m} \frac{2}{\sqrt{f'(r_h) g'(r_h)}} d \tilde{\omega}, \label{imaction2}
\end{equation}
where $r_h = r_h (M-\tilde{\omega}, Q-q)$ and correspondingly $f'(r_h) = f'\left(r_h(M-\tilde{\omega}, Q-q)\right)$
and $g'(r_h) = g'\left(r_h(M-\tilde{\omega}, Q-q)\right)$. In the following,
by taking the tunneling probability as $\Gamma \sim e^{-2 {\rm Im} S}$ and equating it to the Boltzmann factor $e^{-\frac{\omega}{T}}$,
we can obtain the Hawking temperature from eq.~(\ref{imaction2}).

Now making the expansions,
\begin{eqnarray}
f'\left(r_h(M-\tilde{\omega}, Q-q)\right)&=& f'\left(r_h(M, Q-q)\right) - f''\left(r_h(M, Q-q)\right) \frac{\partial r_h}
{\partial M}\bigg{|}_{\tilde{\omega}=0} \ \tilde{\omega}+\mathcal{O}(\tilde{\omega}),\nonumber\\
g'\left(r_h(M-\tilde{\omega}, Q-q)\right)&=& g'\left(r_h(M, Q-q)\right) - g''\left(r_h(M, Q-q)\right) \frac{\partial r_h}
{\partial M}\bigg{|}_{\tilde{\omega}=0} \ \tilde{\omega} + \mathcal{O}(\tilde{\omega}),\label{expansion}
\end{eqnarray}
and substituting them into eq.~(\ref{imaction2}),
we see that only the leading order terms, $f'(r_h(M, Q-q))$ and $g'(r_h(M, Q-q))$, have contributions to the linear term of $\omega$ in ${\rm Im} S$.
As a consequence, the Hawking temperature is determined by $f'(r_h(M, Q-q))$ and $g'(r_h(M, Q-q))$ completely.
Omitting higher order terms in the expansions, we can finish the integral in eq.~(\ref{imaction2}) straightforwardly,
\begin{equation}
{\rm Im} S = \frac{2 \pi}{\sqrt{f'\left(r_h(M, Q-q)\right) g'\left(r_h(M, Q-q)\right)}} (\omega - m).\label{action}
\end{equation}
Therefore, the tunneling probability takes the form,
\begin{equation}
\Gamma \sim e^{- 2 {\rm Im} S} = e^{- \frac{4 \pi}{\sqrt{f'\left(r_h(M, Q-q)\right) g'\left(r_h(M, Q-q)\right)}} (\omega - m)}. \label{tunnelprobability}
\end{equation}
By equating $e^{-2 {\rm Im} S}$ to the Boltzmann factor $e^{-\frac{\omega}{T}}$, we work out the modified Hawking temperature,
\begin{equation}
T_H = \frac{\sqrt{f'\left(r_h(M, Q-q)\right) g'\left(r_h(M, Q-q)\right)}}{4\pi}.\label{hawkingtemperature}
\end{equation}
Note that eq.~(\ref{hawkingtemperature}) in the zeroth order of $q$ recovers the standard Hawking temperature
for a general static spherical charged black hole. In general, the modified Hawking temperature depends on the tunneling particle's charge.

\section{An example: the Reissner-Nordstr\"{o}m black hole}
For the Reissner-Nordstr\"{o}m black hole, we have
\begin{eqnarray}
f(r) &=& g(r) = 1- \frac{2M}{r} + \frac{Q^2}{r^2} = \frac{(r-r_+)(r-r_-)}{r^2},\\
h(r) &=& -\frac{Q}{r},
\end{eqnarray}
where the event horizons $r_\pm$ are
\begin{equation}
r_{\pm} = M \pm \sqrt{M^2 - Q^2}.
\end{equation}
As the tunneling particle crosses the outer horizon, we take $r_h = r_+$. Therefore, the imaginary part of the action (eq.~(\ref{imaction2}))
takes the form,
\begin{equation}
{\rm Im} S = \pi \int^{\omega}_{m} \frac{2 r_+^2}{r_+ - r_-} d \tilde{\omega}.\label{imaction3}
\end{equation}
Considering the self-gravitation of the tunneling particle with energy $\tilde{\omega}$ and charge $q$,
we obtain the event horizons $r_\pm$, and the initial
and final positions of the tunneling particle, $r_{in}$ and $r_{out}$, as follows:
\begin{eqnarray}
r_\pm &=& M - \tilde{\omega} \pm \sqrt{(M - \tilde{\omega})^2 - (Q - q)^2},\label{horizon}\\
r_{in} &=& M + \sqrt{M^2 - Q^2},\label{rin}\\
r_{out} &=& M-\omega + \sqrt{(M-\omega)^2 - (Q-q)^2}.\label{rout}
\end{eqnarray}
Note that\footnote{It is obvious that we have $r_{in}>r_h>r_{out}$ from eqs.~(\ref{horizon}), (\ref{rin}) and (\ref{rout}),
which gives rise to a pole at $r=r_h$ in the integrand of eq.~(\ref{imaction}).
Therefore, following the procedure for deriving eq.~(\ref{imaction2}) from eq.~(\ref{imaction})
we obtain eq.~(\ref{imaction3}) for the Reissner-Nordstr\"{o}m black hole.}
eqs.~(\ref{rin}) and (\ref{rout}) have been used when we derive eq.~(\ref{imaction3}).
Substituting eq.~(\ref{horizon}) into eq.~(\ref{imaction3}) and finishing the integral we have
\begin{eqnarray}
{\rm Im} S &=& \pi \left[(M-m)^2 - (M-\omega)^2 + (M-m)\sqrt{(M-m)^2 - (Q-q)^2}\right.\nonumber\\
& &\left.- (M-\omega)\sqrt{(M-\omega)^2 - (Q-q)^2}\right].\label{RNimaction}
\end{eqnarray}
Eq.~(\ref{RNimaction}) implies that the emission spectrum is not precisely thermal. As a result, the tunneling rate has the form,
\begin{equation}
\Gamma \sim e^{-2 {\rm Im} S} \sim e^{\Delta S_{\rm BH}},\label{entropy}
\end{equation}
where $\Delta S_{\rm BH} = S_{\rm BH} (M-\omega, Q-q) - S_{\rm BH} (M, Q)$
is the difference of the entropy of the black hole after and before the emission.
Expanding ${\rm Im} S$ in $\omega$, $q$ and $m$ to the second order, we obtain,
\begin{eqnarray}
{\rm Im} S &=&\frac{\pi \left(M + \sqrt{M^2-Q^2}\right)^2}{\sqrt{M^2-Q^2}} \left\{\left[1 + \frac{q Q \left(3M^2 - Q^2\right)}{\left(M^2-Q^2\right)
\left(M + \sqrt{M^2-Q^2}\right)^2}\right](\omega - m)\right.\nonumber\\
& &\left.+\frac{\left(M^2-Q^2\right)^{3/2}- M Q^2}{\left(M^2-Q^2\right)\left(M + \sqrt{M^2-Q^2}\right)^2} \left(m^2 - \omega^2\right)
+ \mathcal{O} (\omega^2, m^2, q^2,\omega m,\omega q,mq)\right\}.\label{Im S}
\end{eqnarray}
From the first order term of the above equation with respect to $\omega$, we work out the corresponding modified Hawking temperature,
\begin{eqnarray}
T_H&=&\frac{\sqrt{M^2-Q^2}}{2\pi\left[\left(M+\sqrt{M^2-Q^2}\right)^2+qQ\frac{3M^2-Q^2}{M^2-Q^2}\right]}\nonumber \\
&=&\frac{\sqrt{M^2-Q^2}}{2\pi\left(M+\sqrt{M^2-Q^2}\right)^2}\left[1-\frac{(3M^2-Q^2)Q}{(M^2-Q^2)\left(M+\sqrt{M^2-Q^2}\right)^2} \ q
+\mathcal{O}(q)\right].\label{RNTH}
\end{eqnarray}
In the following we analyze the characteristics of our result from both the charge and mass viewpoints.

If we set $q=0$ in the above equation,
which corresponds to a neutral tunneling particle, we obtain the standard Hawking temperature.
For the Reissner-Nordstr\"{o}m black hole we give the concrete relation between the modified Hawking temperature and
the charge of the tunneling particle.
This relation implies an interesting inequality,  $T_H|_{qQ<0} > T_H|_{q=0} > T_H|_{qQ>0}$,
which shows that the Reissner-Nordstr\"{o}m black hole is more likely to emit particles with the charge opposite to that of the black hole
because a higher temperature corresponds to a larger tunneling rate.
The accumulation of radiation increases the charge of the black hole and finally makes it equal
the mass of the black hole. Consequently, the fate of the Reissner-Nordstr\"{o}m black hole is an extreme black hole rather than a
Schwarzschild black hole if we only consider the effect caused by the Hawking radiation.
This is the way that the charge of the tunneling particle  affects
the Reissner-Nordstr\"{o}m black hole in accordance with the modified Hawking temperature,
which is novel and different from that of ref.~\cite{Zheng Zhao JHEP}.
Incidentally,
%(of course, here we do not consider other mechanics, for example the Schwinger mechanics, which might enforce the black hole to be more likely
%to emit particles with the same sign as the black hole's).
we have made a reasonable assumption $|q/Q|\ll 1$ when we derive the temperature expression eq.~(\ref{RNTH}) from eq.~(\ref{RNimaction}),
that is, the charge of each tunneling particle should be much smaller than that of the black hole.
As a result, the temperature eq.~(\ref{RNTH}) is certainly positive.
%Alternatively, the positivity can easily be seen from the consistency can also be observed from eq.~(\ref{RNimaction}),
%which is valid for arbitrary $q$, that $ImS > 0$ so $\Gamma<1$.

In addition, it seems to be natural that the particle mass appears in the tunneling rate $\Gamma \sim e^{- 2 {\rm Im} S}$.
Because the spacetime tends to the Minkowski type
at infinity, the observer there detects particles only on shell. As the conserved energy $\omega$ can be identified with the energy of
the tunneling particle at infinity, i.e. the dispersion relation is satisfied,  $\omega$ is definitely larger than the mass $m$.
This ensures the consistency of eq.~(\ref {entropy}), which means ${\rm Im} S>0$.
With eq.~(\ref {Im S}) in its first order approximation, we see that the mass produces a shift with respect
to the energy $\om$ on the emission spectrum.
This is also different from that of ref.~\cite{Zheng Zhao JHEP}.
We note that it is not an artifact of the expansion that the mass does not correct the deformed Hawking temperature.
As we can see from eq.~(\ref {RNimaction}) which corresponds to the expression before being expanded,
the introduction of the mass in ${\rm Im} S$ has no influence on the coefficient of $\omega$.
(When $m=0$, eq.~(\ref {RNimaction}) reduces to the formulation associated with
the tunneling of massless particles.) Precisely speaking,
the Hawking temperature is proportional to the inverse of the coefficient of $\omega$ in ${\rm Im}S$ (eq.~(\ref {RNimaction})).
As a result, the mass of the particle has no effects on the modified Hawking temperature. However,
it is not trivial that the mass produces no effects on the modified Hawking temperature but the shift with respect to the energy $\om$
on the emission spectrum.
That is, the non-triviality of the effect of the mass corresponds to the physical explanation that
the particle can be detected only on-shell at infinity.

\section{Conclusion}
In this note, we discuss the Hawking radiation of a general spherical charged black hole by following the idea of ref.~\cite{Wilczek99}
but generalizing it to a quantum tunneling particle with mass and charge.
The mass of the tunneling particle gives rise to a time-like trajectory, and
the trajectory is not geodesic because of the electromagnetic interaction between the charged tunneling particle and black hole.
Furthermore, when we change the momentum integral into the energy integral, see eq.~(\ref{imaction}),
we have to impose a nonzero lower limit of energy integral $m$, see the clarifications under eq.~(\ref{imaction}).
%The reason is that the conserved energy can be identified with the energy of the tunneling particle which is on shell at infinity.
We give the formula of the modified Hawking temperature which depends on the charge of the tunneling particle (see eq.~(\ref{hawkingtemperature})).

Taking the Reissner-Nordstr\"{o}m black hole as an example,
we compute the corresponding tunneling rate and modified Hawking temperature,
and find that our result is different from that of ref.~\cite{Zheng Zhao JHEP}.
In our result both the mass and charge of the tunneling particle have effects on the emission spectrum which
is not precisely thermal either as was first pointed out in ref.~\cite{Wilczek99}.
To the first order approximation, we deduce a charge-modified Hawking temperature (see eq.~(\ref{RNTH})) which recovers the standard Hawking temperature
in the limit of zero charge.
The modified temperature indicates that the Reissner-Nordstr\"{o}m black hole approaches an
extreme black hole under the accumulation of the Hawking radiation effect.
%It makes the standard Hawking temperature seems to be effective because of the dependence of the charge of tunneling particle.
As to the effect caused by the mass of the tunneling particles,
a shift with respect to the energy is produced on the emission spectrum.
As we know, the observer at infinity can detect the tunneling particles only on shell. Consequently,
the emission spectrum is expected to have a lower cutoff compared with the massless emission spectrum.
%Due to our result, the cutoff is a constant related to the mass radiating.

Actually we provide a way to realize the idea of ref.~\cite{Wilczek99} for massive and charged tunneling particles.
We are considering to apply our method to other kinds of black holes, and shall give the results separately.

\vskip 2mm

\section*{Acknowledgments}
Y-GM would like to thank J.-X. Lu of the Interdisciplinary Center for Theoretical Study (ICTS),
University of Science and Technology of China (USTC) for warm hospitality.
This work is supported in part by the National Natural Science Foundation of China under grants No.11175090 and No.10675061.


\newpage

\begin{thebibliography}{99}

%\{Hawking74}
\bibitem{Hawking}
 S.W. Hawking, {\it Black hole explosions}, Nature {\bf 248} (1974) 30;\\
 S.W. Hawking, {\it Particle creation by black holes}, Commun. Math. Phys. {\bf 43} (1975) 199
[Erratum-ibid. {\bf 46} (1976) 206].

 %\{Wilczek99}
\bibitem{Wilczek99}
M.K. Parikh and F. Wilczek, {\it Hawking radiation as tunneling}, Phys. Rev. Lett. {\bf 85} (2000) 5042
[arXiv:hep-th/9907001];\\
M.K. Parikh, {\it A secret tunnel through the horizon}, Int. J. Mod. Phys. {\bf D 13} (2004) 2351; Gen. Rel. Grav. {\bf 36} (2004) 2419
[arXiv:hep-th/0405160].

%\{Zheng Zhao NPB}
%\bibitem{Zheng Zhao NPB}
%Jingyi Zhang and Zheng Zhao, {\it Massive particles' black hole tunneling and de Sitter tunneling}, Nuclear
%Physics {\bf B} {\bf 725} 173-180 (2005);.

%\{Zheng Zhao JHEP}
\bibitem{Zheng Zhao JHEP}
J.-Y. Zhang and Z. Zhao, {\it Hawking radiation of charged
particles via tunneling from the Reissner-Nordstr\"om black hole}, JHEP {\bf 10} (2005) 055.

%{Bin Chen}
%\bibitem{Bin Chen}
%Zhibo Xu and Bin Chen, {\it Hawking Radiation from General Kerr-(anti)de Sitter Black
%Holes}, Phys. Rev. {\bf D} {\bf 75} 024041 (2007) [arXiv:hep-th/0612261].

%{Mehdipour}
%\bibitem{Mehdipour}
%S.H. Mehdipour, {\it Parikh-Wilczek tunneling as massive particles from noncommutative
%Schwarzschild black hole} Commun.Theor.Phys. {\bf 52}:865-870 (2009).

%{Srinivasan}
\bibitem{Srinivasan}
K. Srinivasan and T. Padmanabhan, {\it Particle production and complex path analysis}, Phys.
Rev. {\bf D 60} (1999) 024007 [arXiv:gr-qc/9812028].

\bibitem{Angheben:2005}
M. Angheben, M. Nadalini, L. Vanzo and S. Zerbini, {\it Hawking radiation as tunneling for extremal and rotating black holes},
JHEP {\bf 05} (2005) 014 [arXiv:hep-th/0503081].

%\{kerner}
\bibitem{Kerner}
R. Kerner and R.B. Mann, {\it Tunneling, temperature, and Taub-NUT black holes}, Phys. Rev. {\bf D 73} (2006) 104010 [arXiv:gr-qc/0603019].

%\{Banerjee}
\bibitem{Banerjee}
R. Banerjee and B.R. Majhi, {\it Quantum tunneling beyond semiclassical approximation}, JHEP {\bf 06} (2008) 095 [arXiv:0805.2220[hep-th]].

\bibitem{Painleve:1921}
P. Painlev\'{e}, {\it La m\'{e}canique classique er la thorie de relativit\'{e},} C. R. Acad. Sci. (Paris) {\bf 173} (1921) 677.


\bibitem{SKW}S. Stotyn, K. Schleich and  D. Witt, {\it Observer dependent horizon temperatures: a coordinate-free formulation of Hawking radiation
as tunneling}, Class. Quant. Grav. {\bf 26} (2009) 065010 [arXiv:0809.5093[gr-qc]].

\end{thebibliography}
\end{document}